\documentclass[twocolumn,showpacs,superscriptaddress,aps,prb]{revtex4}
\usepackage{amsmath,amssymb,bm}
\usepackage{graphicx,epsfig} 

\newcommand{\iChEM}{{\it i}{\rm ChEM}}

\newcommand{\itm}{\vspace{0.2 em}\noindent$\bullet\ \, $}

\newcommand{\Opm}{\hat O_{\mbox{\tiny $\pm$}}}

\newcommand{\tS}{\mbox{\tiny S}}
\newcommand{\B}{\mbox{\tiny B}}
\newcommand{\SB}{\mbox{\tiny SB}}
\newcommand{\T}{\mbox{\tiny T}}

\newcommand{\ti}{\Tilde}
\newcommand{\wti}{\widetilde}
\newcommand{\dg}{\dagger}

\newcommand{\ra}{\rangle}

\newcommand{\nl}{\nonumber \\}
\newcommand{\Sec}[1]{Sec.\;\ref{#1}}

\newcommand{\be}{\begin{equation}}
\newcommand{\ee}{\end{equation}}
\newcommand{\bsube}{\begin{subequations}}
\newcommand{\esube}{\end{subequations}}
\newcommand{\Eq}[1]{Eq.\,(\ref{#1})}
\newcommand{\Eqs}[1]{Eqs.\,(\ref{#1})}
\newcommand{\Fig}[1]{Fig.\,\ref{#1}}

\begin{document}

\title{Manipulating quantum coherence of charge states
  in interacting double-dot Aharonov-Bohm interferometers}

\author{Jinshuang Jin} \email{jsjin@hznu.edu.cn}
\affiliation{Department of Physics, Hangzhou Normal University, Hangzhou, Zhejiang 310036, China}

\author{Shikuan Wang}
\affiliation{Department of Physics, Hangzhou Dianzi University, Hangzhou 310018, China}

\author{Jiahuan Zhou}
\affiliation{Department of Physics, Hangzhou Normal University, Hangzhou, Zhejiang 310036, China}

\author{Wei-Min Zhang} \email{wzhang@mail.ncku.edu.tw}
\affiliation{Department of Physics and Center for Quantum
Information Science, National Cheng Kung University, Tainan 70101,
Taiwan}

\author{YiJing Yan} \email{yanyj@ustc.edu.cn}

\affiliation{Hefei National Laboratory for Physical Sciences at the Microscale
and Collaborative Innovation Center of Chemistry for Energy Materials (\iChEM),
University of Science and Technology of China, Hefei 230026, China}

\date{\today}

\begin{abstract}

 We investigate the dynamics of charge--states
coherence in a degenerate double--dot Aharonov--Bohm
interferometer with finite interdot Coulomb interactions.
The quantum coherence of the charge states
is found to be sensitive to the transport setup configurations,
involving both the single--electron impurity channels and the Coulomb--assisted ones.
We numerically demonstrate the emergence of a complete coherence between the two charge states,
with the relative phase being continuously controllable through the magnetic flux.
Remarkably, a fully coherent charge qubit arises
at the double--dots electron pair tunneling resonance condition,
where the chemical potential of one electrode
is tuned at the center between a single--electron impurity channel
and the related Coulomb--assisted channel.
This pure quantum state of charge qubit could be
\emph{experimentally located}
at the current--voltage characteristic turnover position,
where differential conductance sign changes.
We further elaborate the underlying mechanism for both the real--time
and the stationary charge--states coherences in the double--dot systems
of study.

\end{abstract}

\pacs{03.65.Yz, 71.27.+a, 73.23.Hk, 73.63.Kv}
\maketitle

\section{Introduction}

The investigations of semiconductor quantum dots
have long aroused a great deal of attentions.
The various controllability of quantum dots, in terms of not only the geometric shape and size
but also the internal energy levels and couplings,
make it particularly useful in serving as good testbeds for the study of mesoscopic physics,
as well as in potential applications for nanotechnology and quantum information processing.
Especially, quantum dots are the promising candidates for the realizations of scalable quantum computer,
implemented with the electron charge and/or spin qubits.
\cite{Kan98133,Los98120,Han071217,Hay03226804,Fuj06759}
Much progress has been made for the investigation of quantum coherence dynamics,
in particular, the manipulation of charge--state coherence in the lateral
double--dot systems. \cite{Hay03226804,Fuj06759,Tu08235311,Pet10246804,Dov11161302,Cao131401,Kim15243}

On the other hand, quantum coherence transport through parallel double--dots
embedded in Ahronov--Bohm (AB) interferometers
has also been extensively studied both experimentally \cite{Sch97417,Hol01256802,Sig06036804,Hat11076801}
and theoretically.  \cite{Ent02166801,Kon013855,Kon02045316,
Li09521,Har13235426,Tok07113,Szt07386224,Bed14235411,%
Bed12155324,Bed13045418,Rep16165425,Pul10256801,Tu11115318,Tu12115453,Liu16045403}
The particular interest in a ring--structured AB interferometer
is the electron interference which
can be tuned by an externally applied magnetic flux.
The resulted coherent transport property has been
characterized via conductance oscillation
in magnetic flux. \cite{Hol01256802,Sig06036804}
It has been widely studied in quantum transport the relation of
the coherence of AB oscillations to
Coulomb interaction, \cite{Kon013855,Kon02045316,
Li09521,Har13235426,Tok07113,Bed14235411,Szt07386224}
interdot tunneling,\cite{Szt07386224,Hat11076801} and inelastic electron cotunneling
processes, \cite{Sig06036804,Rep16165425} etc.
%
However, the coherent dynamics of the AB double-dot charge states
has not yet been explored in depth when the Coulomb interaction
is fully taken into account because of the theoretical difficulty.
Preliminary studies
demonstrated that the intrinsic dynamics of charge states would
display just a phase localization rather than coherence in a symmetric geometry setup,
\cite{Tu11115318,Bed12155324} unless an asymmetrical geometrical setup is arranged \cite{Tu12115453}
or a large Coulomb interaction is included. \cite{Tok07113,Bed12155324}

 In this work, we study the coherence dynamics of charge states
in double--dot AB interferometers, in the presence of finite
interdot Coulomb interaction.
The analysis is carried out based on the well--established
nonperturbative hierarchical equations--of--motion (HEOM)  approach.\cite{Jin08234703,Li12266403,Ye16608}
The quantum coherence of charge states is shown to be
sensitive to the transport regimes of the electron tunneling channels, including
single--electron impurity channels and Coulomb ones in double dots.
We find that AB double--dots, with finite interdot Coulomb interaction,
 would be very suitable for
the preparation of a fully coherent charge qubit.
The relative phase of the charge qubit
is continuously controllable through the magnetic flux,
rather than the phase localization, as studied previously
on the weak or noninteracting counterparts.\cite{Tu11115318,Bed12155324}
In particular, a fully coherent charge qubit emerges
at the double--dots electron pair tunneling resonance,
when the chemical potential of one electrode
matches with the center between a single--electron impurity channel and the related Coulomb channel.
This pure quantum state of charge quit could be
\emph{experimentally located}
at the current--voltage characteristic turnover position,
where differential conductance changes sign,
from negative  (positive)
to positive (negative).
Finally, using a transformation to reformulate the problem,
we elaborate the underlying mechanism for the real--time  dynamics
of the nonequilibrium charge--states coherence from weak to strong interdot Coulomb interactions.

The rest of paper is organized as follows.
In \Sec{thMet},  we introduce the standard transport model
of the double--dot AB interferometers and
briefly outline the HEOM approach for describing the
coherence dynamics of the charge states in the double-dot.
In \Sec{thcoh}, we present the converged stationary results on the quantum
coherence of the charge qubit in different tunneling regimes,
in the presence of finite interdot Coulomb interaction.
We then study the real--time dynamics and elaborate the
underlying mechanism of the observed nonequilibrium charge-states coherence
in \Sec{thsec4}.
Finally, we give the summary 
in \Sec{thsum}.

\section{Methodology}
\label{thMet}

Consider the nonequilibirum electron transport through a parallel double-dot
embedded in an AB interferometer,
its total Hamiltonian, $H_{\T}=H_{\tS}+H_{\B}+H_{\SB}$,
consists of three parts.
The central parallel double-dot system
is modeled by
\be\label{Hs}
 H_{\tS} =\sum_{u=1,2}\varepsilon_{u}\hat n_u +U \hat n_1\hat n_2,
\ \ \text{with} \ \ \hat n_u=a^{\dg}_{u}a_{u}.
\ee
Here, $a_{u}$ ($a^{\dagger}_{u}$) denotes the annihilation (creation) operator of the electron in
the dot-$u$ orbital state of energy $\varepsilon_{u}$,
and $U$ is the interdot Coulomb interaction.
The electrodes are modeled as noninteracting electrons
reservoirs bath, i.e.,
\be\label{HB}
H_{\B} = \sum_{\alpha k}(\epsilon_{\alpha k}+\mu_{\alpha})c^{\dg}_{\alpha k}
 c_{\alpha k}
 \ee
 with $\alpha=L,R$, under the applied bias voltage potential
$eV=\mu_L-\mu_R$.
Here, $c^{\dg}_{\alpha k}$ ($c_{\alpha k}$)
denotes the creation (annihilation) operator of the electron
with momentum $k$ in the specified $\alpha$-reservoir.
The electrons tunneling between the dots and the reservoirs is
described by the tunneling Hamiltonian,
\be\label{Hsb}
  H_{\SB}=\sum_{\alpha u k} \left( e^{i\phi_{\alpha u}}t_{\alpha u k} a^{\dg}_{u}c_{\alpha k}
    +{\rm H.c.}\right),
\ee
with the AB flux $\Phi$--induced phase factors satisfying
\be\label{AB_phase_sum}
 \phi_{L1}-\phi_{L2}+\phi_{R2}-\phi_{R1}=\phi\equiv
 2\pi\Phi/\Phi_{0}.
\ee
Here, $\Phi_{0}$ denotes the flux quantum.
Without loss of generality,
we adopt (due to the gauge invariant)\cite{Tu11115318,Bed13045418}
\be\label{AB_phases}
  \phi_{L1}=-\phi_{L2}=\phi_{R2}=-\phi_{R1}=\phi/4.
\ee
The hybridization spectral function assumes Lorentzian,
\begin{align}\label{J_Drude}
 J_{\alpha uv}(\omega)
&\equiv \pi e^{i(\phi_{\alpha v}-\phi_{\alpha u})}
 \sum_k  t^\ast_{\alpha u k}t_{\alpha v k}\delta(\omega-\epsilon_{\alpha k})
\nl&
 =\frac{\Gamma_{\alpha uv}W^2}{\omega^2+W^2},
\end{align}
with the equal coupling strengths,
\be\label{AB_Gamma}
\begin{split}
  \Gamma_{\alpha 11}&=\Gamma_{\alpha 22}=\Gamma/2,
\\
 \Gamma^\ast_{L 12}&=\Gamma_{L 21}=\Gamma_{R 12}=\Gamma^{\ast}_{R 21}=e^{ i \phi/2}\Gamma/2.
\end{split}
\ee
Throughout this work, we set the unit of $e=\hbar=1$,
for the electron charge and the Planck constant.
In numerical calculations we set $\mu_L=-\mu_R=V/2$ and
fix the bandwidth at $W=10$\,meV for electrodes.

 In close contact to experiments,\cite{Fuj06759,Han071217}
we set the spinless double--dots to be degenerate,
i.e., $\varepsilon_1=\varepsilon_2=\varepsilon$ in \Eq{Hs}.
The optimized coherence would then be anticipated.
The involved states in the double dots are $|0\ra=|00\ra$, $|1\ra=|10\ra$, $|2\ra=|01\ra$,
and $|d\ra\equiv|11\ra$, i.e., the empty, the dot-$1$
 occupied, the dot-$2$ occupied, and double--dots--occupancy states,
respectively.
The quantum coherence properties of
the double--dots states are described by
the reduced system density matrix,
$\rho(t)\equiv {\rm tr}_{\B}\rho_{\rm tot}(t)$, i.e.,
 the partial trace of the total density operator
$\rho_{\rm tot}$ over the electrode bath degrees of freedom.

  We implement the celebrated HEOM
formalism,\cite{Jin08234703} 
\begin{align}\label{HEOM}
  \dot\rho^{(n)}_{\bf j}(t)&=-\bigg(i{\cal L}_{\tS}
   +\sum_{r=1}^n \gamma_{j_r}\bigg)\rho^{(n)}_{\bf j}(t)
  -i\sum_{j} {\cal A}_{\bar j}\rho^{(n+1)}_{{\bf j}j}(t)   \nl
&\quad
  -i \sum_{r=1}^n (-)^{n-r}{\cal C}_{j_r}\rho^{(n-1)}_{{\bf j}^-_r}(t),
\end{align}
to accurately evaluate the real--time dynamics of the reduced
system density matrix, $\rho^{(0)}(t)\equiv \rho(t)$,
whereas $\rho^{(n)}_{\bf j}(t)\equiv \rho^{(n)}_{j_1\cdots j_n}(t)$,
with $\rho^{(n<0)}(t)\equiv 0$.
In \Eq{HEOM}, ${\cal L}_{\tS}\,\cdot\, \equiv [H_{\tS},\,\cdot\,]$
defines the reduced system Liouvillian;
$j\equiv(\sigma,\alpha,u,\kappa)$ and $\bar j\equiv(\bar\sigma,\alpha,u,\kappa)$
denote the specified collective indexes.
Here, $\sigma =+,-$, and $\bar\sigma$ is its opposite sign;
$\kappa$ arises from the
\emph{nonequilibrium} interacting reservoirs bath correlation
functions,\cite{Jin08234703,Li12266403,Ye16608,Hu10101106,Hu11244106}
in an exponent expansion form
of $C^{\sigma}_{\alpha uv}(t)
 = \sum_{\kappa=1}^{K} \eta^{\sigma}_{\alpha uv\kappa}e^{-\gamma^{\sigma}_{\alpha\kappa}t}$.
Together with denoting $a^+_{u}\equiv a^\dg_{u}$,
and $a^-_{u}\equiv a_{u}$,
the Grassmannian superoperators, ${\cal A}_{\bar j}\equiv {\cal A}^{\bar\sigma}_{\alpha u\kappa}
= {\cal A}^{\bar\sigma}_{u}$
and ${\cal C}_{j}\equiv {\cal C}^{\sigma}_{\alpha u\kappa}$
in \Eq{HEOM}, are defined via\cite{Jin08234703,Li12266403,Ye16608}
\be\label{calAC}
\begin{split}
 {\cal A}^{\sigma}_{u} \Opm &\equiv
    a^{\sigma}_{ u}\Opm \pm \Opm a^{\sigma}_{u}
 \equiv \big[ a^{\sigma}_{u},\Opm\big]_\pm \, ,
\\
 {\cal C}^{\sigma}_{\alpha u\kappa} \Opm  &\equiv
  \sum_{v} \big(\eta^{\sigma}_{\alpha uv\kappa} a^{\sigma}_{v}\Opm
  \mp \eta^{\bar\sigma\,{\ast}}_{\alpha uv\kappa}\Opm a^{\sigma}_{\kappa}\big) .
\end{split}
\ee
Here,
$\Opm$ denotes an arbitrary operator,
with even ($+$) or odd ($-$)  fermionic parity,
such as $\rho^{(2m)}$ or $\rho^{(2m+1)}$, respectively.

   The stationary solutions to HEOM (\ref{HEOM}) can be obtained
by using the conditions, $\{\dot\rho^{(n);{\rm st}}_{\bf j}=0; \forall n\}$.
These together with the normalization constraint, ${\rm tr}\rho^{(0)}=1$,
lead to \Eq{HEOM} a set of coupled linear equations for solving $\{\dot{\rho}^{(n);{\rm st}}_{\bf j}=0\}$.
In practical calculations, an iterative quasiminimal
residual algorithm\cite{Fre91315,Ste07195115} is employed for solving the
large--sized coupled linear equations.\cite{Ye16608}
Equation (\ref{HEOM}) can also be called the
dissipaton equation of motion (DEOM).\cite{Yan14054105,Jin15234108,Yan16110306}
The latter is a quasi--particle theory, which
identities the physical meaning of individual
$\rho^{(n)}_{\bf j}(t)\equiv \rho^{(n)}_{j_1\cdots j_n}(t)$.
Besides \Eq{HEOM}, the DEOM theory includes also the underlying disspaton algebra, especially the
generalized Wick's theorem.\cite{Yan14054105,Jin15234108,Yan16110306}
This extends the real--time dynamics further
to the interacting bath subspace.
Not only the transient transport current,\cite{Zhe08184112,Zhe08093016,Wan13035129}
but also the nonequilibrium current--current
correlation functions can then be evaluated.\cite{Jin15234108}

 As a nonperturbative theory, HEOM
usually converges rapidly and uniformly.\cite{Li12266403,Ye16608,Zhe09164708,Hou15104112}
The hierarchy can be
terminated simply by setting all $\rho^{(n>L)}_{\bf j}=0$,
at a sufficiently large $L$.
%
For the AB double--dot system exemplified in this work,
the HEOM evaluations effectively
converge at the $L=3$ tier level.

\section {Coherence of charge qubit}
\label{thcoh}

\subsection{Coherence control with bias voltage}
\label{thcohA}

  We focus on the quantum coherence of the two charge states,
$\{|1\ra,|2\ra\}$, which constitute a charge qubit,
with the single--electron occupation.
The interested charge--qubit density operator
$\rho_e(t)$ is the $2\times 2$ sub-matrix
of the reduced system $\rho(t)$.  The latter and also
other $\rho^{(n)}_{\bf j}(t)$ spans over
all the four Fock states, $\{|0\ra, |1\ra,|2\ra, |d\ra\}$.
The probability of single electron occupation is given by
$p_{e}={\rm tr}\rho_e = \rho_{11}+\rho_{22}\leq 1$.
The nonzero probabilities of the empty and
double--occupation states, $\rho_{00}$
and $\rho_{dd}$, are the leakage effects. \cite{Tu11115318,Jin13064706}
Denote $\delta = \rho_{11}-\rho_{22}$ for
the probability difference between the two charge states.
The charge qubit entropy is given by $S_e=- {\rm tr}(\varrho_e\ln\varrho_e)$,
with $\varrho_e=\rho_e/{\rm tr}(\rho_e)$ and
$S^{\text{max}}_e = \ln 2$ for two--level systems.
Thus $\chi_e=1-S_e/\ln 2$, satisfying $0\leq \chi_e \leq 1$,
can be used a purity measure,
with $\chi_e=1$ indicating a truly pure state of the charge qubit.

\begin{figure}
\includegraphics[width=1.0\columnwidth]{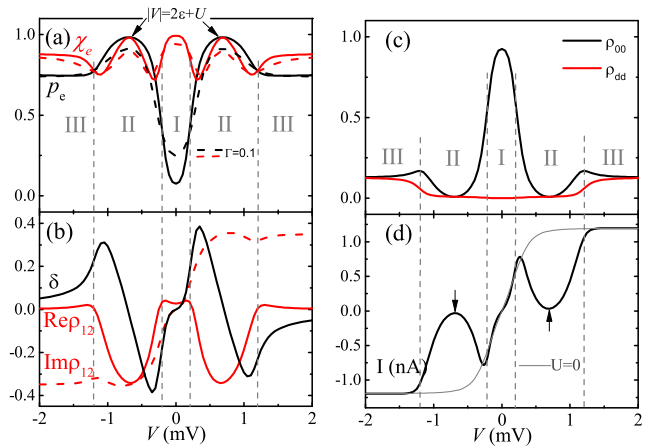}
\caption{ (Color online)
The accurate results for the stationary charge state varying with bias voltage $V$
through different tunneling regimes at the fixed flux $\phi=\pi/2$ and Coulomb interaction $U=0.5{\rm meV}$.
(a) The probability of single-electron occupation and the purity measurement of the resulted
  charge qubit (the solid lines for the weak tunneling rate $\Gamma=0.02{\rm meV}$,
  and the dashed lines for the strong tunneling rate $\Gamma=0.1{\rm meV}$).
(b) The probability difference ($\delta$ ) between the two charge states and their coherence
    term ($\rho_{12}$).
(c) The leakage effects of the probabilities of the empty and the double occupation states.
(d) The average current (the gray-line for noninteracting system $U=0$).
The other parameters are $k_{\B} T=0.02{\rm meV}$,
$\varepsilon=0.1 {\rm meV}$, and $\Gamma=0.02{\rm meV}$ for (b)-(c). }
\label{fig1}
\end{figure}

 Figure \ref{fig1} reports the nonequilibrium
steady--state results, as functions of the applied voltage,
on the charge--quit state properties given in (a) and (b),
the leakage effects in (c),
and the transport current in (d),
respectively, where
the Coulomb interaction of  $U=0.5$\,meV
and the AB phase of $\phi=\pi/2$ are used,
see the description in the figure caption for the details.

  The observed results can be understood
in terms of the interplay between different
tunneling channels involved in individual
transport regimes.
First of all, there are two types of tunneling channels:
The single-electron impurity channels with the degenerate energy levels
at $\varepsilon_1=\varepsilon_2=\varepsilon$,
and the Coulomb--assisted channels at $\varepsilon+U$.
Concerning further their positions in relation to the applied voltage
window, we identify three transport regimes, indicated in \Fig{fig1}
in terms of I, II, and III, respectively.
Let us start with the double--dots state versus
voltage, the $\rho$-$V$ characteristics, reported in
\Fig{fig1}(a)--(c).

\itm
 Regime I: $\varepsilon+U>\varepsilon>\mu_{R(L)}$. This is the
cotunneling regime, with the bias window containing
no tunneling channels.
The resulted single--electron occupation ($p_e=\rho_{11}+\rho_{22}$)
and double occupation ($\rho_{dd}$) are both negligible.
The full probability of empty state ($\rho_{00}$) emerges.

\itm
 Regime II: $\varepsilon+U \!>\!\mu_{L(R)}\!>\!\varepsilon \!>\!\mu_{R(L)}$.
This is the Coulomb--blockade (CB) regime,
and is of particular interest in the present work.
The most striking scenario occurs
at the bias voltage of $|V|=2\varepsilon+U$.
There emerges a nearly pure charge qubit state, with
the single--electron occupation, $p_e=\rho_{11}+\rho_{22}$,
and the purity parameter, $\chi_e$, being both
in close proximity to their maximum values of 1.
These results are specified with the arrows on the solid curves in \Fig{fig1}\,(a),
where $\Gamma=0.02$\,meV and $k_BT=0.02$\,meV are adopted
for demonstration.
At $|V|=2\varepsilon+U$, while $\rho_{00}+\rho_{dd}=1-p_e \simeq 0$,
we have also $\delta=\rho_{11}-\rho_{22}= 0$ and therefore
$|\rho_{12}|\simeq 0.5$, as seen in
\Fig{fig1}\,(b) and (c).
The dashes curves in \Fig{fig1}\,(a) goes with
the increased lead coupling strength, $\Gamma=0.1$\,meV.
Apparently, increasing temperature also
decreases the purity of the charge qubit state.
Note that in the present symmetric bias setup,
$|V|=2\varepsilon+U$ amounts actually to the
pair tunneling resonance condition,\cite{Lei09156803}
which would also occur in the
Coulomb participated regime (II$'$), where
$\mu_{L}>\varepsilon+U>\mu_{R}>\varepsilon$;
see \Sec{thcohB} for the details.

\itm
 Regime III: $\mu_{L(R)}\!>\!\varepsilon,\varepsilon+U \!>\!\mu_{R(L)}$.
This is the sequential-dominated regime, as both the single--electron
impurity and Coulomb--assisted tunneling channels
fall inside the bias window.
The results here are similar to those of $U=0$
and  weak $U$ obtained in Refs.\,\onlinecite{Bed12155324}
and \onlinecite{Tu11115318}, respectively.
Indeed, as reported there before,
the localization of the phase,
$\theta=\arctan[{\rm Im}(\rho_{12})/{\rm Re}(\rho_{12})]$,
appears at the value of $\theta=-\pi/2$ or $\pi/2$.
Either of these two values corresponds to ${\rm Re}\,\rho_{12}=0$ [cf.\ \Fig{fig1}\,(b)].
The fact that ${\rm Re}\,\rho_{12}$ vanishes
in the sequential-dominated regime (III)
is rather robust against the flux; see the discussion for \Fig{fig2} later.

 Figure \ref{fig1}\,(d) depicts the current-voltage ($I$-$V$) characteristics.
Particularly, in the CB regime (II) that does not exist
for noninteracting ($U=0$; grey--curve) case,
the $I$-$V$ curve exhibits
a remarkable concave down (or up) feature,
for $V>0$ (or $V<0$).
The turnover positions, located with the arrows in \Fig{fig1}\,(d),
are right at $|V|=2\varepsilon+U$.
In other words, the nearly pure
charge qubit state, with both the purity
and occupation number parameters,
$\chi_e$ and $p_e$ [see \Fig{fig1}\,(a)], in close proximity to their maximum values of 1,
could be \emph{experimentally located}
at the aforementioned $I$-$V$ characteristic turnover position,
where the differential conductance sign changes.

\subsection{Charge qubit phase at pair transfer resonance
 versus Aharonov-Bohm magnetic flux
}
\label{thcohB}

 Examine now $\theta=\arctan[{\rm Im}(\rho_{12})/{\rm Re}(\rho_{12})]$,
the charge qubit phase, as functions of the magnetic flux.
We focus on the case of $V=|2\varepsilon+U|$, at which
the charge qubit is in close proximity to
a pure state; cf.\ \Fig{fig1}(a).
It is noticed that in the present symmetric bias
setup, $V=|2\varepsilon+U|$ satisfies the pair tunneling resonance condition,\cite{Lei09156803}
\be\label{PTR}
  \mu_{\alpha}-\varepsilon=\varepsilon+U-\mu_{\alpha}.
\ee
Beside the CB regime (II) described earlier,
the pair tunneling occurs also in another
transport setup configuration:
\\
\itm Regime II$'$: $\mu_L>\varepsilon+U>\mu_R>\varepsilon$,
the Coulomb participation (CP) regime.
In this regime, the transport would be
primarily driven by
the electron tunneling from dots to $R$-lead.
The relevant pair tunneling resonance is therefore
\Eq{PTR} with $\alpha=R$.
On the other hand, in the CB regime (II), where $\varepsilon+U>\mu_L>\varepsilon>\mu_R$,
the resonance follows \Eq{PTR} with $\alpha=L$,
as the transport would now be primarily driven by
the electron tunneling from $L$-lead to dots.

\begin{figure}
\includegraphics[width=1.\columnwidth]{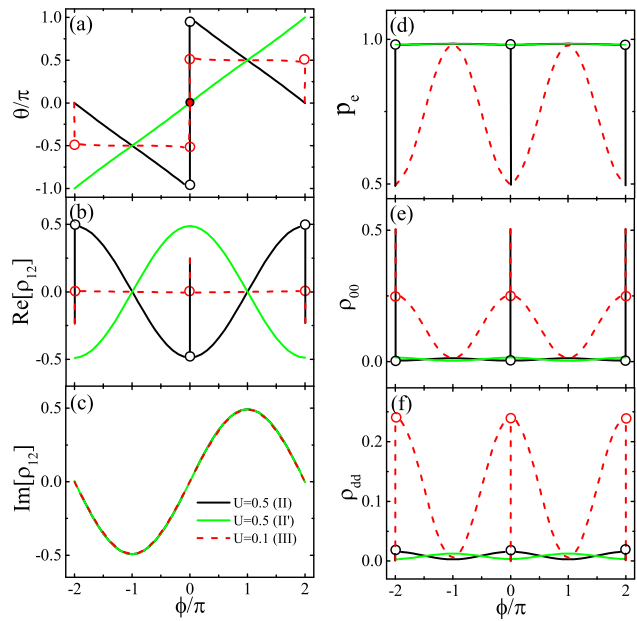} 
\caption{ (Color online)
The accurate results for the stationary charge state
varies with the magnetic flux for different Coulomb interaction at the bias voltage $V=0.7 {\rm meV}$.
(a) The characteristic coherence is denoted by the relative phase $\theta$,
 and (b) and (c) the real and imaginary parts of $\rho_{12}$.
The probabilities of the states for single-electron occupation, empty, and double occupation
are illustrated in (d), (e), and (f), respectively.
 Here, $U=0.1$\,meV is for sequential-dominated regime (III), and $U=0.5$\,meV is for
 CB regime (II) (black--curves, where $\varepsilon=0.1$\,meV)
 and CP regime (II$'$) (green--curves, where $\varepsilon=-0.6$\,meV)
 with satisfying $V=|2\varepsilon+U|$.
 The other parameters are the same as in \Fig{fig1}.
}
\label{fig2}
\end{figure}

 Figure \ref{fig2} reports the nonequilibrium
steady--state results, as functions of
the applied AB magnetic flux, in terms
of $\phi/\pi \equiv 2\Phi/\Phi_0$.
Both the CB (black--curves)
and CP (green--curves) cases
are operated at the pair tunneling resonance
voltage, $V=|2\varepsilon+U|=0.7$\,meV,
with the same $U=0.5$\,meV,
but different values of $\varepsilon=0.1$\,meV and $-0.6$\,meV,
respectively.
Included for comparison are also the sequential-dominated regime (III) counterparts (red--dashed--curves),
exemplified with $U=0.1$\,meV and $\varepsilon=0.1$\,meV at $V=0.7$\,meV.
As that in \Fig{fig1}, where $\phi=\pi/2$,
the sequential-dominated regime displays
the phase localization at $\theta=-\pi/2$ or $\pi/2$,
for $V<0$ and $V>0$, respectively.\cite{Tu11115318}
This result is independent of the flux at
$\phi\neq 2m\pi$, with $m$ being an integer, as shown by
the red--dashed horizontal parts in \Fig{fig2}\,(a),
where Re\,$\rho_{12}=0$, see \Fig{fig2}\,(b).
The single--electron occupation, $p_e$, and the leakage effects,
$\rho_{00}$ and $\rho_{dd}$,
as shown by the red--dashed curves in \Fig{fig2}\,(d)--(f),
also agree with those
noninteracting ($U=0$) results reported in Ref.\ \onlinecite{Tu11115318}.

On the other hand, in both the CB (II) and CP (II$'$) regimes,
while ${\rm Im}\rho_{12}$ is rather independent of
the interdot Coulomb coupling,
${\rm Re}\rho_{12}$ significantly deviates
from the zero-value behavior in the $U=0$ case.
The single--electron occupation is remarkably enhanced,
and meanwhile the leakage effects are greatly suppressed.
%
Especially,  at $|V|=|2\varepsilon+U|$ that satisfies the pair tunneling resonance
condition,\cite{Lei09156803}
the charge qubit state,
as inferred from \Fig{fig1}(a) and (b)
for its $\chi_e\simeq 1$
and $\delta\equiv\rho_{11}-\rho_{22}=0$,
assumes a pure--state proximity of
\be\label{psi}
 |\psi\ra=\frac{1}{\sqrt{2}}\big(|1\ra+ e^{-i\theta}|2\ra \big).
\ee
The AB magnetic flux--tuned phase, as shown in  \Fig{fig2}\,(a)  for $V>0$,
is given by
\be\label{theta1}
{\theta}= \begin{cases}
  \pi - \phi/2;  & \text{CB regime (II)}
\\
 \phi/2; &  \text{CP regime (II$'$)}
\end{cases} .
\ee
For the bias $V<0$, the above relations hold with exchange of $\phi$ to $-\phi$
(not shown in \Fig{fig2}),
due to the phase--lead symmetry relations underlying \Eq{AB_phases}.

 Remarkably,  as shown in \Fig{fig2}, while the singularity occurs
in the CB regime (II) at $\phi = 2m\pi$,
the CP counterparts are completely free of the singularity.
The nearly pure charge qubit state, with both the purity parameter, $\chi_e$,
and occupation number, $p_e$, in close proximity to their maximum values of 1.
It is interesting to notice that
in the present CP regime (II$'$) setup, the double--occupation
level locates below the transport window; i.e., $\epsilon_d=2\varepsilon+U=-0.7\,\text{meV}<\mu_R=-0.35\,\text{meV}$
in study.
However, its occupation number $\rho_{dd}$ remains very small,
under the pair tunneling resonance voltage; see \Fig{fig2}\,(f).
Involved here is also the \emph{interference resonance}
that overcomes the leakage from the desired charge qubit state.


\section{Coherence dynamics analysis}
\label{thsec4}

\subsection{Charge qubit coherence dynamics}
\label{thsec4A}

\begin{figure}
\includegraphics[width=1.0\columnwidth]{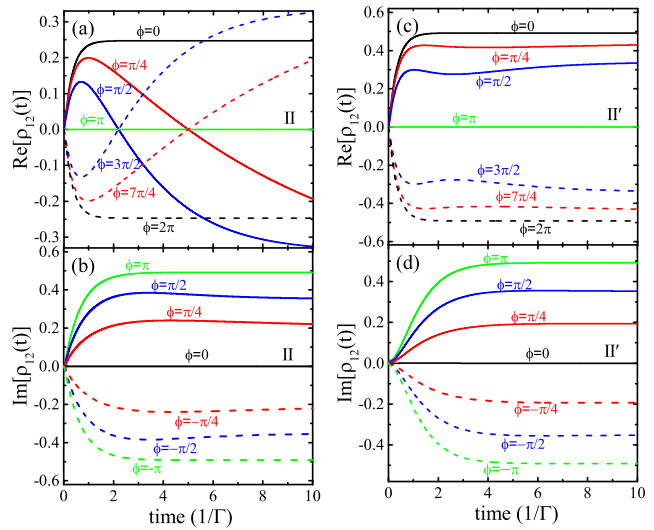}
\caption{ (Color online)
The coherence evolution of $\rho_{12}(t)$ for CB regime (II) ($\varepsilon=0.1$\,meV) and
the CP regime (II$'$) ($\varepsilon=-0.6$\,meV) setups in
the left (a and b) and right (c and d) panels, respectively,
with the strong interdot Coulomb interaction $U=0.5$\,meV.
The other parameters are the same as in \Fig{fig2}.
}
\label{fig3}
\end{figure}

To further explore the underlying machnism of the full coherence realization of the
charge qubit states in the AB interferometers, we study the evolution
of $\rho_{12}(t)$, the transient charge--qubit coherence,
in both the CB regime (II) (left--panels) and the CP regime (II$'$)
(right--panels), with the initial empty state in the double dots ($\rho_{00}(0)=1$).
The results are presented in  \Fig{fig3}.
%
It shows that the short-time ($t \lesssim 1/\Gamma$) dynamics
in both the regime II and II$'$ are quite similar
as that in the sequential-dominated regime (III) reported previously in Refs.\,\onlinecite{Bed12155324} and \onlinecite{Tu11115318}.
%
The short-time dynamics of $\rho_{12}(t)$ is dominated by the electron tunneling through
the single--electron impurity channels ($\varepsilon$),
with little contributions from the Coulomb--assisted channel ($\varepsilon+U$).
Denote $\rho_{12}(t)=|\rho_{12}(t)|e^{i\theta(t)}$,
the relative phase between the two charge states, $|1\ra$ and $|2\ra$,
is found to be $\theta(t)=\phi/2$ when $t \lesssim1/\Gamma$.
For $t> 1/\Gamma$, the charge--qubit coherence becomes
sensitive to the specific tunneling regimes.
The relative locations of the single--occupation ($\varepsilon$) and
the double--occupation ($\varepsilon+U$) transport channels
with respect to the bias window
take the crucial role when $t> 1/\Gamma$.
Especially the nonequilibrium charge qubit
operated in either the CB regime (II) or the CP regime (II$'$)
remains a proximity to a full coherence dynamics,
as shown by \Eq{psi} in the long--time limit.
The characteristics of Im\,$\rho_{12}(t)$ are similar in these two transport regimes,
without changing signs as the time evolves, while
Re\,$\rho_{12}(t)$ behaviors differ remarkably.
In the CB regime (II) (see \Fig{fig3}\,(a)),
Re\,$\rho_{12}(t)$ will switch the sign (unless $\phi= 2m\pi$ that is associated with $\theta=\pi/2$
for $V>0$), which
leads to the relative phase $\theta=\pi-\phi/2$.
On the other hand, in the CP regime (II$'$), as shown in \Fig{fig3}\,(b),
Re\,$\rho_{12}(t)$ has no such sign change and the
relative phase remains as the case of $\theta(t)=\phi/2$.
Note that the above results in the CB regime and the CP regime
could be changed significantly by weakening the Coulomb interaction
or increasing the bias voltage or temperature.
For example, the full coherence of charge qubit dynamics would be immediately
broken down and the charge qubit state is reduced to
the phase localization,
similar to the noninteracting ($U=0$) limit in the previous studies.\cite{Tu11115318,Bed12155324}

\subsection{Mechanistic analysis}
\label{thsec4B}

\begin{figure}
\includegraphics[width=0.70\columnwidth]{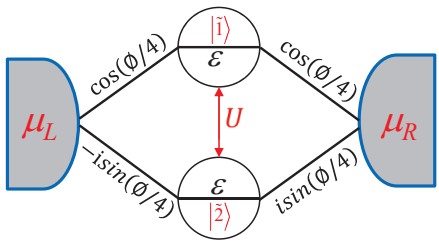} 
\caption{(Color online)
 The schematic view of a double-dot AB interferometer in
 the new basis in terms of the transformation \Eq{unit}
 and the corresponding tunneling Hamiltonian of \Eq{Hsb2}.
}
\label{fig4}
\end{figure}

The above nonequilibirum features on the
charge--states coherence dynamics,
including the short--time, long--time and stationary state behaviors,
can be understood as follows.
Taking the following transformation on the electron operators in the dots,
\be\label{unit}
 \ti a_{1} = ({a}_1+{a}_2)/{\sqrt{2}}\ \ \text{and}
\ \ \ti a_{2} = ({ a}_1-{ a}_2)/{\sqrt{2}}.
\ee
The system and bath Hamiltonians, \Eqs{Hs} and (\ref{HB}),
are invariant under this transformation.
The tunneling Hamiltonian, \Eq{Hsb}, becomes
\be\label{Hsb2}
 {\wti H}_{\SB}=\sum_{\alpha u k} \left( {\ti t}_{\alpha u k}  {\ti a}^{\dg}_{u}c_{\alpha k}
    +{\rm H.c.}\right),
\ee
with ${\ti t}_{\alpha 1 k} =\sqrt{2}\,t_{\alpha 1 k}\cos(\phi/4)$
and ${\ti t}^{\ast}_{L 2 k}={\ti t}_{R 2 k}
 = i\sqrt{2}\,t_{\alpha 2 k}\sin(\phi/4)$.
The above results are schematically depicted
in \Fig{fig4}.
The transformed hybridization spectral function,
with the equal coupling strengthes of \Eq{AB_Gamma},
remain the Lorentzian form of \Eq{J_Drude}, but having
 \be\label{tigamma}
\begin{split}
\wti\Gamma_{\alpha 11}&=2\Gamma\cos^2(\phi/4),
\ \ \
\wti\Gamma_{\alpha 22} =2\Gamma\sin^2(\phi/4),
\\
\wti\Gamma_{L 21}&=\wti\Gamma^\ast_{L 12} =\wti\Gamma_{R 12} =\wti\Gamma^{\ast}_{R 21}
 =i\,\Gamma\sin(\phi/2).
\end{split}
\ee
From $|u\ra = a^{\dg}_u|0\ra$
and $|\ti u\ra = \ti a^{\dg}_u|0\ra$ with
\Eq{unit},
we have
\be\label{rhounit}
\begin{split}
\rho_{11}&= \left(\rho_{\ti 1\ti 1}+\rho_{\ti 2\ti 2}+2\,{\rm Re}\,\rho_{\ti 1\ti 2}\right)/2,
\\
\rho_{22}&= \left(\rho_{\ti 1\ti 1}+\rho_{\ti 2\ti 2}-2\,{\rm Re}\,\rho_{\ti 1\ti 2}\right)/2,
\\
\rho_{12}&= \left(\rho_{\ti 1\ti 1}-\rho_{\ti 2\ti 2}-2i\,{\rm Im}\,\rho_{\ti 1\ti 2}\right)/2.
\end{split}
\ee
Equations (\ref{tigamma}) and  (\ref{rhounit}) are used
in the following analysis, with the focus on the flux $\phi$--dependent
charge qubit phase,
$\theta=\arctan[{\rm Im}(\rho_{12})/{\rm Re}(\rho_{12})]$.

 Let us start with the two special scenarios,
where the interdot Coulomb interaction does not play
the role, and the charge qubit phases are
always localized.

 (\emph{i}) At $\phi=2m\pi$, we have $\theta=0$ or $\pi$ (cf.\ the black--lines in \Fig{fig3}).
As inferred from \Eq{tigamma},
this scenario has either $\wti\Gamma_{\alpha 11}=0$ or
$\wti\Gamma_{\alpha 22}=0$,
with odd or even $m$, respectively, and also $\wti\Gamma_{\alpha 12}=0$.
The electron tunnels through only one of the
transformed single--occupation
channels, either $|\ti 2\ra$ or $|\ti 1\ra$.
Consequently, $\rho_{\ti 1\ti 2}=0$, since
there is no interference between
these two states.
In this case, $\rho_{12}(t)$ is always real, as inferred
from \Eq{rhounit}, and the charge qubit phase is localized
at $\theta=0$ or $\pi$.
Physically the above scenario amounts to the transport setup
involving only one single spinless electronic level.
The interdot Coulomb interaction does not play
any roles in this scenario,
and the double--occupation is always $p_d(t)=0$.
The long--time probabilities of the empty
and the single-electron occupied states are equal, i.e.,
$p_{0}=0.5$ and $p_{e}=0.5$.
The latter is via either $\rho_{\ti 1\ti 1}$ or $\rho_{\ti 2\ti 2}$,
exclusively.

(\emph{ii}) At $\phi=(2m+1)\pi$, we have $\theta=\pm\pi/2$
(cf.\ the green--lines in \Fig{fig3}).
As inferred from \Eq{tigamma},
this scenario goes by
$\wti\Gamma_{\alpha 11}=\wti\Gamma_{\alpha 22}=|\wti\Gamma_{\alpha 12}|=\Gamma/2$,
resulting in $\rho_{\ti 1 \ti 1}(t)=\rho_{\ti 2\ti 2}(t)=|\rho_{\ti 1 \ti 2}(t)|$.
This is the case of a \emph{full interference with equal probability}.
In this case, $\rho_{12}(t)$ is always pure imaginary, as inferred
from \Eq{rhounit}, and the charge qubit phase is localized
at $\theta=\pm\pi/2$.
Physically, a full interference with equal probability
is an \emph{interference resonance}.
It leads to the maximum value of $p_e =\rho_{11}+\rho_{22}\approx 1$ in
the long--time region. Both the vacancy and double occupations are suppressed.
This interference resonance behavior is
independent of interdot Coulomb interaction.

 Turn to the situations of $\phi\neq n\pi$,
away from the above two special scenarios,
and the interdot Coulomb interaction will play the roles.
The general remarks on the nonspecial situations are as follows.
(\emph{a})
In the short--time ($t>1/\Gamma$) region,
 electrons tunnel mainly through  two single-electron impurity channels,
with $\varepsilon_{\ti 1}=\varepsilon_{\ti 2}=\varepsilon$.
According to the flux--dependent tunneling rate of \Eq{tigamma},
one of them could be called the fast channel and the other be the slow one.\cite{Saf03136801}
More precisely, $|\ti 1\ra$ is the fast channel when $\phi<\pi/4$,
whereas it is the slow one when $\phi>\pi/4$.
The fast channel dominates in short time.
The above analysis also dictates
the sign of ${\rm Re}[\rho_{12}(t)] = [\rho_{\ti 1\ti 1}(t)-\rho_{\ti 2\ti 2}(t)]/2$
[cf.\ \Eq{rhounit}] in the short time region.
As time evolves, the slow channel occupation
gradually accumulates.
The sign of ${\rm Re}[\rho_{12}(t)]$ would change
if the population inversion could occur.
For example, the individual curve in \Fig{fig3}(a) changes sign,
while that in \Fig{fig3}(c) does not.
We will elaborate these observations later;
(\emph{b}) When $t>1/\Gamma$,
Coulomb--assisted ($\varepsilon+U$)-channels play roles.
These are the transfer channels, rather than
the double--occupation state of energy $2\varepsilon+U$.

 Focus hereafter the long--time behavior for $\phi\neq n\pi$,
which depends on both single--electron $\varepsilon$--channel and
Coulomb--assisted ($\varepsilon+U$)--channel.
Apparently, the nonequilibrium property manifests the
interplay between these two transfer channels
and their relative locations with respect to the bias window.
The cotunneling regime (I) is not the interest
of this work, since it generates no
significant population in the charge qubit state.
On the other hand, the sequential--dominant regime (III),
where $\mu_L>\varepsilon,\varepsilon+U>\mu_R$,
the interested transfer channels both fall inside the bias window.
This is similar to the well--studied
Coulomb--free scenario,\cite{Bed12155324,Tu11115318}
with the results being summarized as follows.
In the wide--band--reservoirs limit,
the probabilities of electrons tunneling through
$|\ti 1\ra$ and $|\ti 2\ra$ would be equal, i.e.,
$\rho_{\ti1\ti1}=\rho_{\ti2\ti2}$
(unless $\phi=2m\pi$, the special scenario-(\emph{i})
described earlier, with $\theta=0$ or $\pi$).
Again, as inferred from \Eq{rhounit},
the resultant $\rho_{12}$ is pure imaginary.
Phase localization occurs at $\theta = \pm\pi/2$,
the same value of the special scenario-(\emph{ii}),
but without the aforementioned full interference resonance condition.
The leakage effect can not be neglected;
see the regime-III parts of \Fig{fig1}\,(c).

 The main contribution of this work
is concerned with the CB regime (II), where $\varepsilon+U>\mu_L>\varepsilon>\mu_R$,
and the CP regime (II$'$) where $\mu_L>\varepsilon+U>\mu_R>\varepsilon$.
The chemical potential of one reservoir falls in between
the two interested tunneling channels.
The electron pair tunneling mechanism is anticipated.\cite{Lei09156803}
Appears at the pair tunneling resonance of \Eq{PTR}
an almost perfect charge qubit, as discussed in detail in \Sec{thcohB}.
The different behaviours, as depicted in \Fig{fig2} and \Fig{fig3}
and also \Eq{theta1}, are rooted at the facts
that in the CB regime it is the single--electron $\varepsilon$--channel
inside the bias window,
whereas in the CP regime it is the Coulomb-assisted $(\varepsilon+U)$--channel.
Actually, the aforementioned fast versus slow $\varepsilon$--channels,
discussed in relation to the short--time region properties,
are physically concerned with the CB regime.
Involves there the \emph{dynamical} Coulomb blockade
processes,\cite{Saf03136801} leading to electron accumulation in the slow channel,
and further the population inversion along evolution.
Consequently, Re\,$\rho_{12}(t)$ experiences the sign change,
as depicted in \Fig{fig3}\,(a).
This also leads to the charge qubit phase transition
at around $\phi=0$; see the black--curve in \Fig{fig2}\,(a).
The CP regime is just the opposite to the CB regime.
Now it is the Coulomb-assisted $\varepsilon+U$--channels inside
the bias window, whereas the single--electron ones are outside.
There are no dynamical Coulomb blockage effects;
neither the slow channel accumulation
nor the population inversion.
The resulted relative phase follows
$\theta=\phi/2$ without jump;
see the green--line in \Fig{fig2}\,(a).

\section{summary}
\label{thsum}

 We have demonstrated that
interdot Coulomb interaction would play a crucial role
in operating a degenerate double--dots as a charge qubit.
Finite Coulomb interaction could result in
dynamical Coulomb--assisted transport channels.
Together with the single--electron ones they
comprise electron tunneling inference pairs,
whenever $\varepsilon+U>\mu_{L}>\varepsilon>\mu_{R}$ (Coulomb blockage
regime)
or $\mu_{L}>\varepsilon+U>\mu_{R}>\varepsilon$
(Coulomb participation without blockage).
The pair tunneling interference is
responsible for the coherence control of a charge qubit,
including its relative phase, via the applied bias voltage
and magnetic flux.
 A fully coherent charge qubit emerges
at the double--dots electron pair tunneling resonance,
$\varepsilon+U-\mu_{\alpha}=\mu_{\alpha}-\varepsilon$
[cf.\ \Eq{PTR}].
This amounts to $|V|=|2\varepsilon+U|$, provided $\mu_L=-\mu_R=V/2$
that the bias voltage is applied symmetrically to two leads.
Interestingly, the pair tunneling resonance can be located
at the $I$-$V$ characteristic turnover position,
as specified by the arrows in \Fig{fig1}(d).
Therefore, the information on a fully coherent charge qubit
would be experimentally extracted from where
the differential conductance sign changes.

 Moreover, the charge qubit phase, operated especially in
the Coulomb participation regime, can be smoothly manipulated via the applied magnetic flux
[cf.\ \Eq{theta1}].
This is different from the Coulomb blockage scenario,
where the Coulomb--assisted $(\varepsilon+U)$--channel is above the
transport window. The underlying dynamical blockage
induces population inversion in the long--time region,
and consequently the relative phase change, as inferred
from \Fig{fig3}(a) and (b).
In contrast, in the Coulomb participation regime,
the $(\varepsilon+U)$--channel is within the
transport window and does not have the aforementioned blockage effect,
as seen from \Fig{fig3}(c) and (d).
All these observations
are elaborated via the real--time dynamical and stationary properties
of the charge qubit coherence,
and also on the basis of a canonical transformation;
see \Sec{thsec4}.

 In summary, we present an experimentally viable 
approach to the preparation and manipulation of charge qubit coherence 
in double--dots systems.
The predictions of this work and the underlying principles
are closely related to the field of quantum information/computation in general.

\acknowledgments
Support from the Natural Science Foundation of China
(Nos. 11675048,11447006 \& 21633006),
the Ministry of Science and Technology of China
(No. 2016YFA0400904), and the
MST of Taiwan  (No.~MST-105-2112-M-006-008-MY3)  
is gratefully acknowledged.


\begin{thebibliography}{10}

\bibitem{Kan98133}
B.~E. Kane,
\newblock Nature {\bf 393}, 133 (1998).

\bibitem{Los98120}
D.~Loss and D.~P. DiVincenzo,
\newblock Phys. Rev. A {\bf 57}, 120 (1998).

\bibitem{Han071217}
R.~Hanson, L.~P. Kouwenhoven, J.~R. Petta, S.~Tarucha, and L.~M.~K.
  Vandersypen,
\newblock Rev. Mod. Phys. {\bf 79}, 1217 (2007).

\bibitem{Hay03226804}
T.~Hayashi, H.~D.~C. T.~Fujisawa, and Y.~Hirayam,
\newblock Phys. Rev. Lett. {\bf 91}, 226804 (2003).

\bibitem{Fuj06759}
T.~Fujisawa, T.~Hayashi, and S.~Sasaki,
\newblock Rep. Prog. Phys {\bf 69}, 759 (2006).

\bibitem{Tu08235311}
M.~W.~Y. Tu and W.-M. Zhang,
\newblock Phys. Rev. B {\bf 78}, 235311 (2008).

\bibitem{Pet10246804}
K.~D. Petersson, J.~R. Petta, H.~Lu, and A.~C. Gossard,
\newblock Phys. Rev. Lett. {\bf 105}, 246804 (2010).

\bibitem{Dov11161302}
Y.~Dovzhenko et~al.,
\newblock Phys. Rev. B {\bf 84}, 161302 (2011).

\bibitem{Cao131401}
G.~Cao et~al.,
\newblock Nat. Commun. {\bf 4}, 1401 (2013).

\bibitem{Kim15243}
D.~Kim et~al.,
\newblock Nat. Nanotechnol. {\bf 10}, 243 (2015).

\bibitem{Sch97417}
R.~Schuster et~al.,
\newblock Nature {\bf 385}, 417 (1997).

\bibitem{Hol01256802}
A.~W. Holleitner, C.~R. Decker, H.~Qin, K.~Eberl, and R.~H. Blick,
\newblock Phys. Rev. Lett. {\bf 87}, 256802 (2001).

\bibitem{Sig06036804}
M.~Sigrist et~al.,
\newblock Phys. Rev. Lett. {\bf 96}, 036804 (2006).

\bibitem{Hat11076801}
T.~Hatano et~al.,
\newblock Phys. Rev. Lett. {\bf 106}, 076801 (2011).

\bibitem{Ent02166801}
O.~Entin-Wohlman, A.~Aharony, Y.~Imry, Y.~Levinson, and A.~Schiller,
\newblock Phys. Rev. Lett. {\bf 88}, 166801 (2002).

\bibitem{Kon013855}
J.~{K\"{o}nig} and Y.~Gefen,
\newblock Phys. Rev. Lett. {\bf 86}, 3855 (2001).

\bibitem{Kon02045316}
J.~K\"onig and Y.~Gefen,
\newblock Phys. Rev. B {\bf 65}, 045316 (2002).

\bibitem{Li09521}
F.~Li, H.~J. Jiao, H.~Wang, J.~Y. Luo, and X.~Q. Li,
\newblock Physica E: Low-dimensional Systems and Nanostructures {\bf 41}, 521
  (2009).

\bibitem{Har13235426}
R.~H\"artle, G.~Cohen, D.~R. Reichman, and A.~J. Millis,
\newblock Phys. Rev. B {\bf 88}, 235426 (2013).

\bibitem{Tok07113}
Y.~Tokura, H.~Nakano, and T.~Kubo,
\newblock New J. Phys. {\bf 9}, 113 (2007).

\bibitem{Szt07386224}
D.~Sztenkiel and R.~\'{S}wirkowicz,
\newblock J. Physics: Condensed Matter {\bf 19}, 386224 (2007).

\bibitem{Bed14235411}
S.~Bedkihal and D.~Segal,
\newblock Phys. Rev. B {\bf 90}, 235411 (2014).

\bibitem{Bed12155324}
S.~Bedkihal and D.~Segal,
\newblock Phys. Rev. B {\bf 85}, 155324 (2012).

\bibitem{Bed13045418}
S.~Bedkihal, M.~Bandyopadhyay, and D.~Segal,
\newblock Phys. Rev. B {\bf 87}, 045418 (2013).

\bibitem{Rep16165425}
E.~V. Repin and I.~S. Burmistrov,
\newblock Phys. Rev. B {\bf 93}, 165425 (2016).

\bibitem{Pul10256801}
V.~I. Puller and Y.~Meir,
\newblock Phys. Rev. Lett. {\bf 104}, 256801 (2010).

\bibitem{Tu11115318}
M.~W.-Y. Tu, W.-M. Zhang, and J.~S. Jin,
\newblock Phys. Rev. B {\bf 83}, 115318 (2011).

\bibitem{Tu12115453}
M.~W.-Y. Tu, W.-M. Zhang, J.~S. Jin, O.~Entin-Wohlman, and A.~Aharony,
\newblock Phys. Rev. B {\bf 86}, 115453 (2012).

\bibitem{Liu16045403}
J.-H. Liu, M.~W.-Y. Tu, and W.-M. Zhang,
\newblock Phys. Rev. B {\bf 94}, 045403 (2016).

\bibitem{Jin08234703}
J.~S. Jin, X.~Zheng, and Y.~J. Yan,
\newblock J. Chem. Phys. {\bf 128}, 234703 (2008).

\bibitem{Li12266403}
Z.~H. Li et~al.,
\newblock Phys. Rev. Lett. {\bf 109}, 266403 (2012).

\bibitem{Ye16608}
L.~Z. Ye et~al.,
\newblock WIREs Comp. Mol. Sci. {\bf 6}, 608–638 (2016).

\bibitem{Hu10101106}
J.~Hu, R.~X. Xu, and Y.~J. Yan,
\newblock J. Chem. Phys. {\bf 133}, 101106 (2010).

\bibitem{Hu11244106}
J.~Hu, M.~Luo, F.~Jiang, R.~X. Xu, and Y.~J. Yan,
\newblock J. Chem. Phys. {\bf 134}, 244106 (2011).

\bibitem{Fre91315}
R.~W. Freund and N.~M. Nachtigal,
\newblock SIAM J. Numer. Math. {\bf 60}, 315 (1991).

\bibitem{Ste07195115}
G.~Stefanucci,
\newblock Phys. Rev. B {\bf 75}, 195115 (2007).

\bibitem{Yan14054105}
Y.~J. Yan,
\newblock J. Chem. Phys. {\bf 140}, 054105 (2014).

\bibitem{Jin15234108}
J.~S. Jin, S.~K. Wang, X.~Zheng, and Y.~J. Yan,
\newblock J. Chem. Phys. {\bf 142}, 234108 (2015).

\bibitem{Yan16110306}
Y.~J. Yan, J.~S. Jin, R.~X. Xu, and X.~Zheng,
\newblock Frontiers Phys. {\bf 11}, 110306 (2016).

\bibitem{Zhe08184112}
X.~Zheng, J.~S. Jin, and Y.~J. Yan,
\newblock J. Chem. Phys. {\bf 129}, 184112 (2008).

\bibitem{Zhe08093016}
X.~Zheng, J.~S. Jin, and Y.~J. Yan,
\newblock New J. Phys. {\bf 10}, 093016 (2008).

\bibitem{Wan13035129}
S.~K. Wang, X.~Zheng, J.~S. Jin, and Y.~J. Yan,
\newblock Phys. Rev. B {\bf 88}, 035129 (2013).

\bibitem{Zhe09164708}
X.~Zheng, J.~S. Jin, S.~Welack, M.~Luo, and Y.~J. Yan,
\newblock J. Chem. Phys. {\bf 130}, 164708 (2009).

\bibitem{Hou15104112}
D.~Hou et~al.,
\newblock J. Chem. Phys. {\bf 142}, 104112 (2015).

\bibitem{Jin13064706}
J.~S. Jin, M.~W.-Y. Tu, N.-E. Wang, and W.-M. Zhang,
\newblock J. Chem. Phys. {\bf 139}, 064706 (2013).

\bibitem{Saf03136801}
S.~S. Safonov et~al.,
\newblock Phys. Rev. Lett. {\bf 91}, 136801 (2003).

\bibitem{Lei09156803}
M.~Leijnse, M.~R. Wegewijs, and M.~H. Hettler,
\newblock Phys. Rev. Lett. {\bf 103}, 156803 (2009).

\bibitem{Bar06246804}
P.~Barthold, F.~Hohls, N.~Maire, K.~Pierz, and R.~J. Haug,
\newblock Phys. Rev. Lett. {\bf 96}, 246804 (2006).

\bibitem{Jin11053704}
J.~S. Jin, X.~Q. Li, M.~Luo, and Y.~J. Yan,
\newblock J. Appl. Phys. {\bf 109}, 053704 (2011).

\bibitem{Jin16083038}
J.~S.~Jin, C.~Karlewski, and M.~Marthaler,
\newblock New J. Phys. {\bf 18}, 083038 (2016).

\bibitem{Gur986602}
S.~A. Gurvitz,
\newblock Phys. Rev. B {\bf 57}, 6602 (1998).

\bibitem{Gur05205341}
S.~A. Gurvitz, D.~Mozyrsky, and G.~P. Berman,
\newblock Phys. Rev. B {\bf 72}, 205341 (2005).

\bibitem{Li05205304}
X.~Q. Li, J.~Y. Luo, Y.~G. Yang, P.~Cui, and Y.~J. Yan,
\newblock Phys. Rev. B {\bf 71}, 205304 (2005).

\end{thebibliography}
\end{document}